\documentstyle[twoside,fleqn,espcrc2,epsfig]{article}

\title{Jet Cross Sections in $\gamma^*\gamma$-Scattering
  at $e^+e^-$ Colliders in NLO QCD}

\author{B.\ P\"otter\address{II. Institut f\"ur Theoretische 
    Physik, Universit\"at Hamburg, Luruper Chaussee 149, \\ D-22761
    Hamburg, Germany (e-mail: poetter@mail.desy.de)}\thanks{Sup\-ported 
    by the Bundes\-ministerium f\"ur For\-schung und
    Tech\-nologie, Bonn, Germany, under Contract 05~7~HH~92P~(0),
    and by EEC Program {\it Human Capital and Mobility} through Network
    {\it Physics at High Energy Colliders} under Contract
    CHRX--CT93--0357 (DG12 COMA).}}

\begin{document}

\begin{abstract}
  Recent results from NLO QCD calculations for inclusive
  jet cross sections in $\gamma^*\gamma$-scattering at 
  $e^+e^-$ colliders, especially for LEP, are reported. The
  virtuality $Q^2$ of the virtual photon is non-zero and can be
  unlimited large. The virtuality of the second photon is zero and the
  spectrum is calculated with the Weizs\"acker-Williams
  approximation. Four components of the cross sections have to be
  distinguished, involving direct and resolved real and virtual photon
  contributions. Since $Q^2$ is non-zero, the virtual photon structure
  function is needed to calculate the contributions involving a
  resolved virtual photon.
\end{abstract}

\maketitle

\section{Introduction}

Jet production in $\gamma\gamma$-scattering is an interesting field,
both to study perturbative QCD and to obtain information about the
partonic structure of the photon. Jet production from two photons can be
obtained at $e^+e^-$-colliders as a subprocess to the reaction 
$e^+e^- \to  e^+e^- + X$. The leptons both radiate photons that, in
the simplest case, couple to a quark-antiquark pair which 
produces jets with high $E_T$ in the final
state. In the case that both leptons disappear undetected in the beam
pipe the photons are quasi-real. This case has been studied at
TRISTAN \cite{01} and LEP \cite{02} experiments. The presence of a large
scale ($E_T$) allows perturbative QCD calculations. These have been
performed \cite{1,2,3,3b} and the agreement of the predictions with
experimental data is good (see e.g. \cite{4}). Recently it has become
possible to detect one of the leptons to perform a single-tag
experiment \cite{05}. The detection of the lepton allows to
reconstruct the virtuality ($Q^2$) of the radiated photon and it is
thus possible to obtain data on jet production for the case of
scattering of virtual on real photons.

It is well-known that the real photon can not only couple directly
to the charge of the bare quarks but can also fluctuate into a bound
state und thus serve as a source of quarks and gluons. This resolved
component is described by a parton distribution function (PDF) of the
real photon. Thus, the scattering of photons also provides a
possibility to study the parton contents of the photon. The resolved
component of the virtual photon deviates from that of the real photon
in that it has a $Q^2$-dependence. At large $Q^2$ the resolved
component of the virtual photon is believed to be negligable. Since
so far only limited data exist for the structure of the virtual
photon \cite{11}, the modeling of the $Q^2$-behaviour of the
virtual photon PDF is still rather ambiguous. Mainly two LO
parametrizations of the virtual photon PDF exist, namely those
of Gl\"uck, Reya and Stratmann \cite{grs}, and of Schuler and
Sj\"ostrand \cite{sas}.

Cross sections have been calculated for $\gamma^*\gamma$-scattering at
LO accuracy some time ago \cite{10}, but LO calculations suffer from rather
large scale and scheme dependences. Thus, it is desirable to perform these
calculations in NLO, which has been achieved recently \cite{10b}.

\section{Partonic Cross Section in NLO QCD}

\begin{figure}[bbb]
  \unitlength1mm
  \begin{picture}(122,36)
    \put(-10,5){\psfig{file=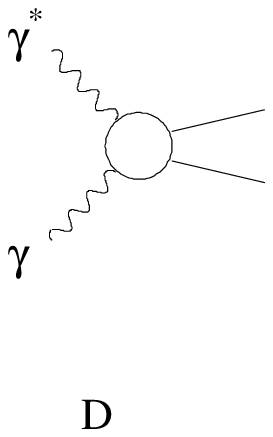,width=6cm}}
    \put(30,5){\psfig{file=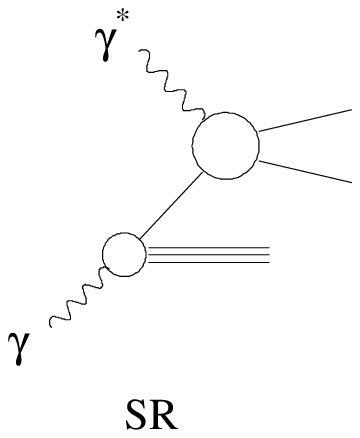,width=6cm}}
    \put(-10,-22){\psfig{file=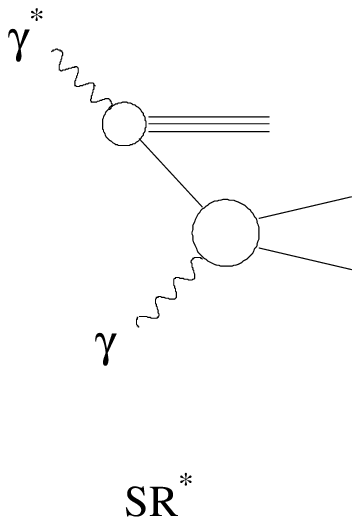,width=6cm}}
    \put(30,-22){\psfig{file=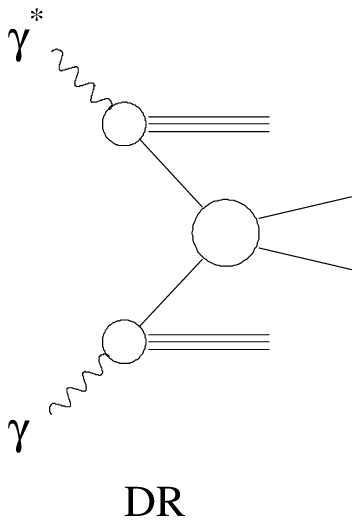,width=6cm}}
  \end{picture}
  \caption{The four different interaction modes in
    $\gamma^*\gamma$-scattering.}
\end{figure}

We will consider $e^+e^-$-scattering cross sections with the
$\gamma^*\gamma$-scattering subprocess
\begin{equation}
  \gamma^*_a(Q^2) + \gamma_b(P^2=0) \to \mbox{jet}_1 + \mbox{jet}_2 +
  \mbox{X} 
\end{equation}
The interaction of a virtual with a real photon can happen in four
different ways, depending on whether the photon interacts directly or
resolved, which is depicted in Fig.~1. The variables $y_a,y_b$
in the following describe the momentum fraction of the photon $a,b$ in
the electron and $x_a,x_b$ describes the momentum fraction of the
parton in the photon $a,b$.

Taking into account both the transverse and longitudinal polarizations
of the virtual photon, the cross section $d\sigma_{e^+e^-}$ for
$e^+e^-$-scattering is conveniently written as the convolution
\begin{eqnarray} 
  \frac{d\sigma_{e^+e^-}}{dQ^2dy_ady_b} = \sum_{a,b} \int dx_adx_b
  F_{\gamma /e^-}(y_b) f_{b/\gamma}(x_b)  \nonumber \\
  \times \frac{\alpha}{2\pi Q^2} \left[ \frac{1+(1-y_a)^2}{y_a}
  f^U_{a/\gamma^*}(x_a)d\sigma_{ab} \right. \nonumber \\ 
  + \left. \frac{2(1-y_a)}{y_a} f^L_{a/\gamma^*}(x_a)d\sigma_{ab}
  \right] \label{e+e-}
\end{eqnarray}
The PDF's of the real and the virtual photon are $f_{b/\gamma}(x_b)$ and 
$f^{U,L}_{a/\gamma^*}(x_a)$, respectively, where $U$ and $L$ denote
the unpolarized and longitudinally polarized photon contributions,
respectively. The direct photon interactions are included in formula
(\ref{e+e-}) through delta functions. For the direct virtual photon one has 
the relation $f^{U,L}_{a/\gamma^*}d\sigma_{ab} = \delta
(1-x_a)d\sigma^{U,L}_{\gamma^*b}$, whereas for the direct real photon
the relation is $f_{a/\gamma}d\sigma_{ab} = \delta
(1-x_b)d\sigma_{\gamma b}$, where $d\sigma_{ab}$ refers to the
partonic cross section. The function $F_{\gamma/e^-}(y_b)$ describes
the spectrum of the real photons emitted from the electron according
to the Weizs\"acker-Williams approximation \cite{wwill}.

The partonic cross sections in LO consist of two final state
particles. The NLO corrections consist of the virtual and real
corrections, which both exhibit characteristic divergencies.
The real corrections for the different subprocesses D, SR,
SR$^*$ and DR have been calculated with the phase-space slicing method
and are available in the literature in NLO \cite{3,10b,graudenz}. 
The sum of real and virtual corrections is finite after factorization
of singularities from the initial state. Of special importance here is
the $\gamma^*\to q\bar{q}$ splitting term. In the limit of
the $q\bar{q}$-pair being collinear the logarithm
\begin{equation}
  M = \ln\left( 1 + \frac{y_ss}{zQ^2}\right) P_{q\leftarrow\gamma}(z) 
\end{equation}
(where $y_s$ is the phase-space-slicing parameter and $s$ is the
partonic cms energy)
can be factorized from the cross section, which is singular for
$Q^2=0$. This singularity is absorbed into the PDF of the virtual
photon with virtuality $Q^2$ in such a way that the
$\overline{\mbox{MS}}$ factorization result of the real photon is
obtained in the limit $Q^2\to 0$ \cite{23}.

\section{Results for Jet Cross Sections}

The different subprocesses D, SR, SR$^*$ and DR have been implemented
into the computer program {\tt JetViP} \cite{jv} which allows to
calculate jet cross sections in $\gamma^*\gamma$-scattering using the
Snowmass jet algorithm \cite{21}. 

\begin{figure}[hhh]
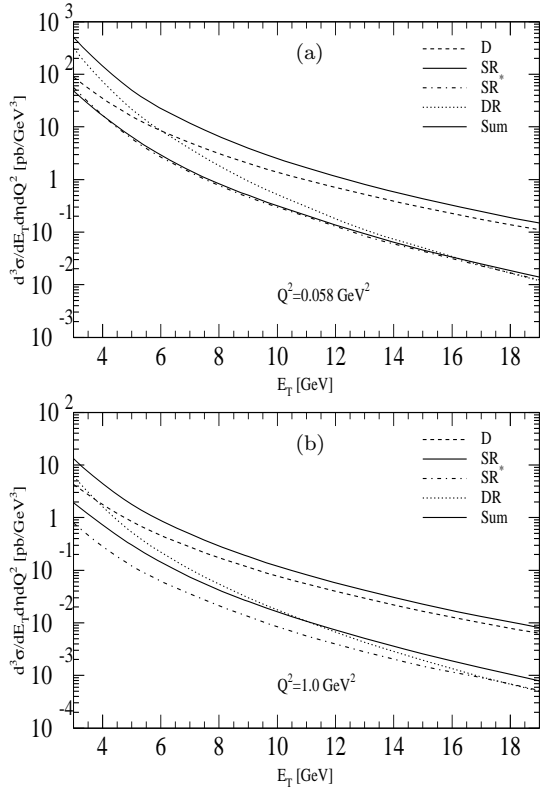

  \unitlength1mm
  \begin{picture}(122,95)
    \put(-2,-16){\epsfig{file=plots/fig.10a,width=8.1cm,height=12cm}}
    \put(-2,-68){\epsfig{file=plots/fig.10c,width=8.1cm,height=12cm}}
    \put(37,87){\footnotesize (a)}
    \put(37,35){\footnotesize (b)}
  \end{picture}
  \caption{Single-jet inclusive cross section integrated over $\eta
    \in [-2,2]$. The upper full curve is the sum of the D,
    SR, SR$^*$ and the DR components. (a) $Q^2=0.058$ GeV$^2$; 
    (b) $Q^2=1.0$ GeV$^2$.}
  \label{10}
\end{figure}

For producing our plots we assume kinematical conditions that
will be encountered at LEP2, where the photons are emitted by
colliding electrons and positrons, both having the energy of
$E_e=83.25$ GeV. We choose the configuration, where the virtual photon
travels in the positive $z$-direction. We consider only the
$\overline{\mbox{MS}}$-GRS \cite{grs} parametrization of the photon
PDF here. We use the PDF of GRS for both, the real and the virtual
photon. We have set the number of flavors to $N_f=4$, adding the
contributions from photon-gluon fusion by fixed order perturbation
theory. The renormalization and factorization scales are set equal, with 
$\mu_R=M_\gamma =M_{\gamma^*}=E_T$.

In Fig.~\ref{10} a and b the $E_T$ spectra for the virtualities
$Q^2=0.058$ and $1.0$ GeV$^2$ for the cross section
$d^3\sigma /dE_Td\eta dQ^2$ are shown, integrated over the interval
$-2\le \eta \le 2$. The value $Q^2_{eff}=0.058$ GeV$^2$
is chosen as to reproduce the $Q^2\simeq 0$ case. The
SR (lower full) and SR$^*$ (dash-dotted) curves coincide in Fig.\
\ref{10} a, where the real photon is approximated by the integrated
Weizs\"acker-Williams formula and the virtual photon has the value
of $Q^2_{eff}$. The full cross section (upper full curve) is dominated
by the DR component in the small $E_T$ range for the small $Q^2$
value. For $Q^2=1.0$ GeV$^2$, the DR and D contributions are of the
same order around $E_T=4$ GeV, but the DR component falls off quickly
for the higher $E_T$'s, leaving the D component as the dominant
contribution. This is expected, since the point-like coupling of the
photons is more important for larger $E_T$ and $Q^2$. Since the
virtual photon contribution is suppressed for larger $Q^2$
the SR$^*$ contribution falls below  the SR curve when going to higher
values of $Q^2$. In all curves, both SR contributions do not play an
important role for the full cross section. Of course, all
contributions decrease with increasing $Q^2$.

\begin{figure*}[ttt]
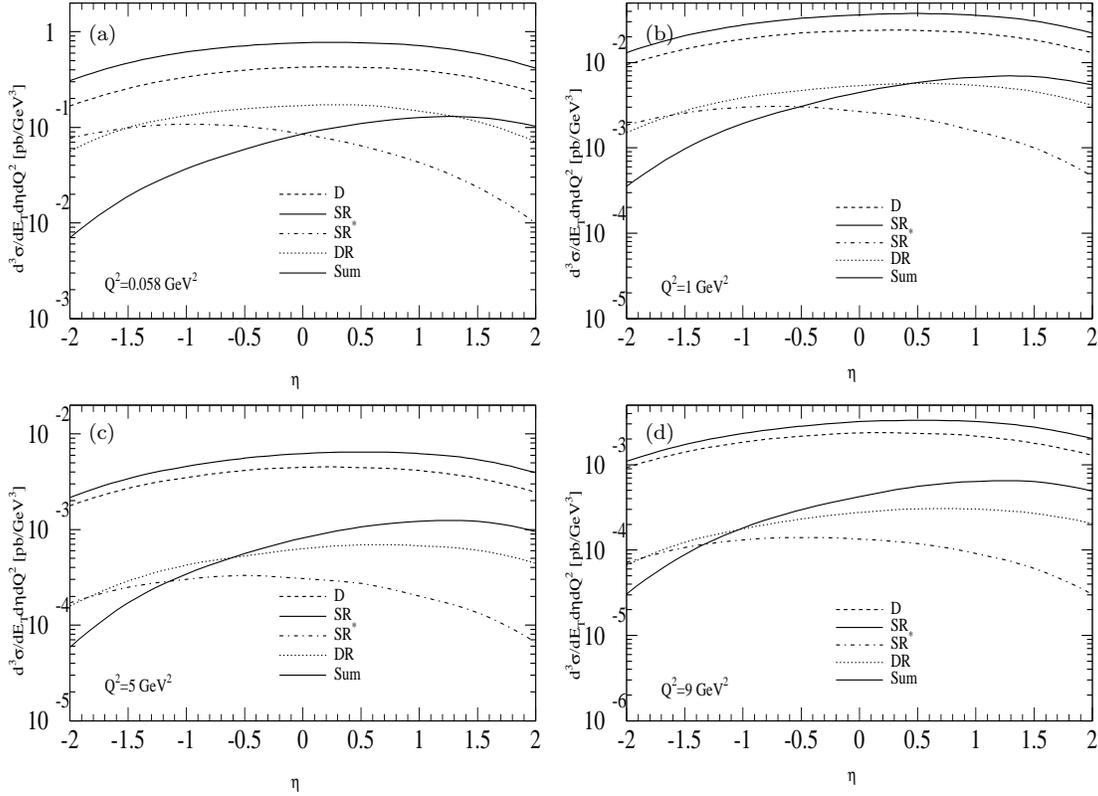

  \unitlength1mm
  \begin{picture}(122,90)
    \put(4,-12){\epsfig{file=plots/fig.11a,width=8.1cm,height=12cm}}
    \put(78,-12){\epsfig{file=plots/fig.11b,width=8.1cm,height=12cm}}
    \put(4,-65.5){\epsfig{file=plots/fig.11c,width=8.1cm,height=12cm}}
    \put(78,-65.5){\epsfig{file=plots/fig.11d,width=8.1cm,height=12cm}}
    \put(16,91){\footnotesize (a)}
    \put(90,91){\footnotesize (b)}
    \put(16,38){\footnotesize (c)}
    \put(90,38){\footnotesize (d)}
  \end{picture}
  \caption{Single-jet inclusive cross section as a function of $\eta$
    for fixed $E_T=10$ GeV. The upper full curve is
    the sum of the D, SR, SR$^*$ and the DR components.
    (a) $Q^2=0.058$ GeV$^2$; (b) $Q^2=0.5$ GeV$^2$; (c) $Q^2=1.0$
    GeV$^2$; (d) $Q^2=9.0$ GeV$^2$.}
  \label{11}
\end{figure*}

We turn to the $\eta$-distribution of the single-jet cross section for
fixed $E_T=10$ GeV between $-2\le \eta \le 2$ for the virtualities
$Q^2=0.058, 1, 5$ and $9$ GeV$^2$, plotted in Fig.~\ref{11} a--d.
The D and DR distributions for the lowest virtuality
$Q^2_{eff}$ are almost symmetric, because of the identical energies of
the incoming leptons. The SR curve falls off for negative $\eta$,
whereas the SR$^*$ component is suppressed for positive $\eta$. Going
to higher $Q^2$ values, the D contribution stays more or less
symmetric and dominates the full cross section, as we have already
seen in Fig.~\ref{10} for the larger $E_T$ values. The components
containing contributions from the resolved virtual photon DR and
SR$^*$ fall of in the region of negative $\eta$ so that they become
more and more asymmetric. This is clear,  since we have chosen the
virtual photon to be incoming from the positive $z$-direction and the
resolved virtual photon contribution is decreasing for higher
virtualities. The DR and SR contributions are of the same magnitude in
the negative $\eta$ region and the DR component is dominant for the
larger $\eta$ values, where the resolved photon is more important. The
same holds for the D and SR$^*$ distributions in the negative $\eta$
region, only here the D component is far more dominant than the SR$^*$
component in the whole $\eta$ region.

\subsection*{Acknowledgements}

I am grateful to G.~Kramer for interesting discussions. 


\end{document}